\documentclass[twocolumn,aps,prc,showpacs,preprintnumbers,amsmath,amssymb]{revtex4}
\usepackage{epsfig} 
\usepackage{dcolumn}

\begin{document}
 
\title{ Comment on ``Electromagnetic dissociation of $^8$B and the astrophysical S-factor for $^7$Be(p,$\gamma$)$^8$B"}

\author{  K. A. Snover}
\altaffiliation{electronic address: snover@u.washington.edu}
\author{A. R. Junghans}
\altaffiliation{present address: Forschungszentrum Rossendorf, Postfach 510119, 01324 Dresden, Germany}
\altaffiliation{electronic address: A.Junghans@fz-rossendorf.de}
\author{E. C. Mohrmann}
\altaffiliation{electronic address: mohrmann@u.washington.edu}
\affiliation{Center for Experimental Nuclear Physics and Astrophysics, University of  Washington,
Seattle,~Washington~98195}

\date{\today}

\begin{abstract}

Recently, Davids and Typel recommended a ``low" value of S$_{17}$(0) based on fits to published direct and Coulomb dissociation data, in which they excluded the precise result of Junghans et al.  We show that their statistical analysis is incorrect, due to a substantial underestimate of the experimental uncertainties, and leads to conclusions that are not supported by a proper analysis.

\end{abstract}

\maketitle

Davids and Typel~\cite{dt} (here called DT) determined S$_{17}$(0) values by fitting their potential model calculations to published low energy ($E \leq$ 400 keV) $^7$Be(p,$\gamma$)$^8$B data from 5 direct and 2 Coulomb dissociation (CD) experiments.  DT found a low probability P($\chi^2, \nu$) = 0.008 for the mean value of S$_{17}$(0) from this data set, and rejected the most precise value, the 2002 measurement by Junghans et al.~\cite{junghans02}, on the basis of its large deviation from the mean.  From a fit to their data set excluding Junghans, they obtained a ``recommended" value S$_{17}$(0) = 18.6 $\pm$ 0.4(exp) $\pm$ 1.1(extrap) eV b, where ``extrap" is their estimated extrapolation uncertainty.  However, DT substantially underestimated the experimental uncertainties in S$_{17}$(0) from the various experiments.  

Table~\ref{data} compares the ``$^7$Be + p potential" fits given in Table II of DT,  with the analysis of low energy data from the same experiments as given in Table VII and Fig. 20 of ref.~\cite{junghans03} (here called JU03).  The top 6 entries in our Table~\ref{data} are from the modern direct experiments~\cite{filippone,hammache,strieder,baby,junghans02, junghans03}, and the bottom 4 from the modern CD experiments~\cite{kikuchi,iwasa,schumann,dt} (see also \cite{motobayashi}).  Since the Descouvement and Baye (DB) theory used by JU03, and the DT potential calculations are very similar in the low energy region (see Fig. 4 of DT), the JU03 and DT fit results should also be very similar~\cite{fitdifference}.  While the central values shown in Table~\ref{data} are indeed similar (the DT results are a few percent lower than the JU03 results), the DT uncertainties are significantly smaller (by factors of 1.3 to 3.1) than the JU03 uncertainties~\cite{dtparameters}.  Several of these uncertainties may also be compared directly with the uncertainties quoted by the original authors.  The S$_{17}$(0) uncertainties quoted in Junghans~\cite{junghans02} and Baby~\cite{baby} from fits to low energy data agree with the uncertainties shown in column 3 of Table~\ref{data} and disagree with the uncertainties shown in column 4.  In fact, all but 2 of the DT uncertainties shown in Table~\ref{data} are consistent with an (incorrect) procedure in which, for each experiment, the systematic uncertainty in the normalization was included with the statistical uncertainty on each fitted data point, instead of being excluded while minimizing $\chi^2$ and then folded into the fit uncertainty.  As a result, the DT analysis erroneously reduced the systematic error, for these experiments, by approximately the square root of the number of fitted data points.  The 2 exceptions are the DT uncertainties from their fits to the data of Filippone and Sch\"umann, which are consistent with no included systematic uncertainty.

\begin{table}
\caption{S$_{17}$(0) values and experimental uncertainties, in eV b, from the low energy DB fits of JU03~\protect\cite{junghans03} and the low energy ``$^7$Be+p potential" fits of DT~\protect\cite{dt}.}
\label{data}
\begin{ruledtabular}
\begin{tabular}{lcll}
First author & Expt. &JU03 fit\footnotemark[1] &  DT fit\footnotemark[2]  \\
    \hline
Filippone & direct & 20.7 $\pm$ 2.5 & 19.8 $\pm$ 0.8  \\
Hammache  & " & 20.1 $\pm$ 1.3 & 19.5 $\pm$ 0.8 \\
Strieder & " & 18.8 $\pm$ 1.8 & 18.7 $\pm$ 1.4  \\ 
Baby & " & 20.8 $\pm$ 1.3 & 20.3 $\pm$ 1.0  \\ 
Junghans 2002 & " & 22.3 $\pm$ 0.7 & 22.2 $\pm$ 0.3  \\
Junghans 2003 & " & 22.1 $\pm$ 0.6\footnotemark[3] & (not fitted)  \\
\hline
Kikuchi & CD & 19.1 $\pm$ 1.7 & (not fitted) \\ 
Iwasa & " & 20.6 $\pm$ 1.2 & (not fitted)  \\ 
Sch\"umann & " & 19.1 $\pm$ 1.5 & 19.0 $\pm$ 1.0  \\ 
Davids & " & 17.4 $\pm$ 1.4 & 17.1 $\pm$ 1.1 \\  
\end{tabular}
\end{ruledtabular}
\footnotetext[1]{from fits to data with $E \leq$ 425 keV.}
\footnotetext[2]{from fits to data with $E \leq$ 400 keV.}
\footnotetext[3]{this value supercedes the Junghans 2002 value.}
\end{table}

\begin{figure}
\includegraphics[width=0.5\textwidth]{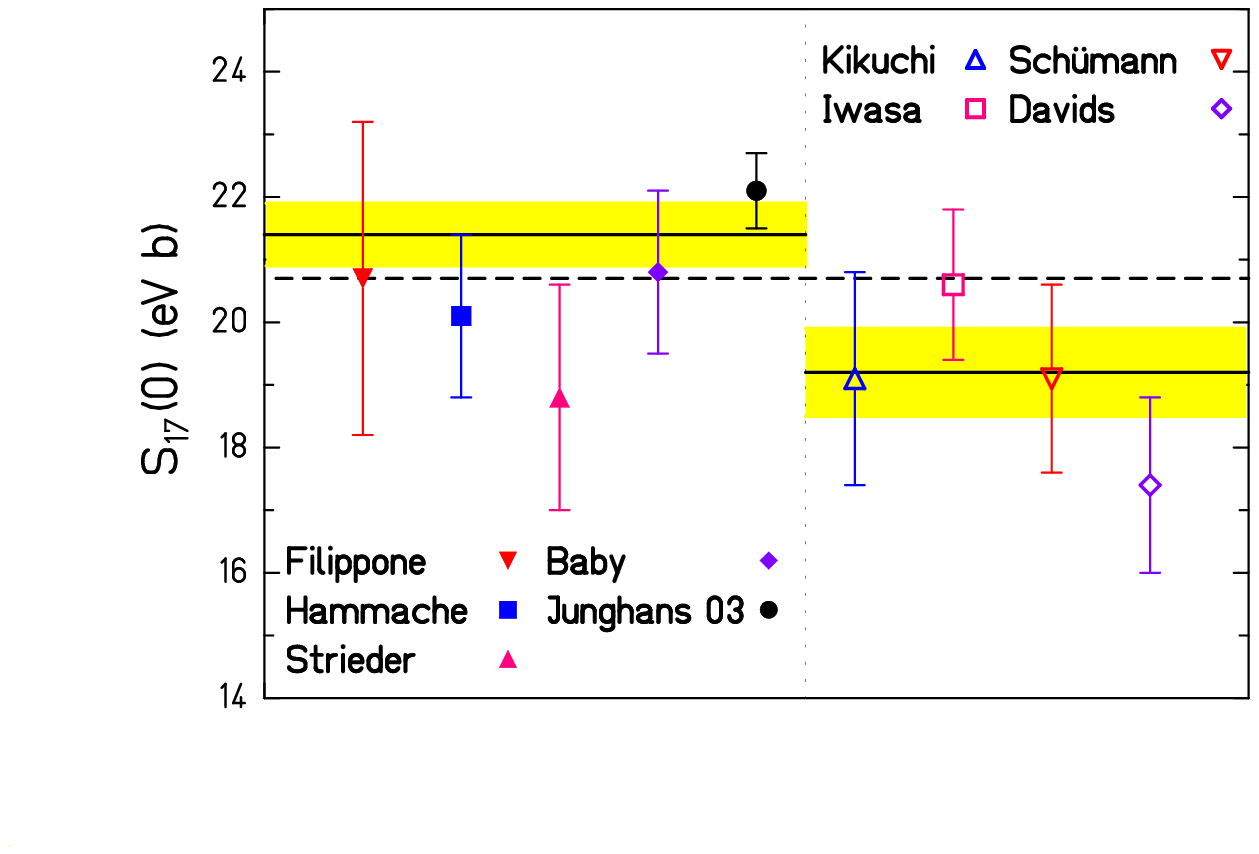}
\caption{(Color online) S$_{17}$(0) values and 1$\sigma$ experimental uncertainties from DB fits given in JU03 and shown in Table~\protect\ref{data}. The Seattle-TRIUMF value is taken from ref.~\protect\cite{junghans03}.  The dashed horizontal line  is the 20.7 eV b mean value from a fit to the full data set.  The solid horizontal lines and shaded bands show the 21.4 $\pm$ 0.5 eV b and 19.2 $\pm$ 0.7 eV b mean values from fits to direct and to CD data, respectively.}
\label{ourS17fit}
\end{figure}

Table~\ref{data} also shows the JU03 fit results for the modern CD experiments of Kikuchi et al.~\cite{kikuchi} and Iwasa et al.~\cite{iwasa}, which in our opinion should be included in comparisons.  Fig.~\ref{ourS17fit} shows the JU03 fit results from Table~\ref{data}~\cite{02vs03}, 
together with the mean value of 20.7 $\pm$ 0.4 eV b~\cite{expterrors} from a fit to the full data set, for which $\chi^2/\nu$ = 1.8 ($\nu$ = 8) and P($\chi^2, \nu$)  = 0.07 (if we exclude the values from Kikuchi and Iwasa, to more closely approximate the conditions of the DT comparison, we obtain 20.8 $\pm$ 0.4 eV b, $\chi^2/\nu$ = 2.2 ($\nu$ = 6) and P($\chi^2, \nu$)  = 0.04).  Neither of these fit probabilities is unacceptably small.  Furthermore, both the Davids and the Junghans points differ from the mean by nearly the same amount (2.3$\sigma$).  Excluding $\it{either}$ Davids $\it{or}$ Junghans results in a good fit, with P($\chi^2, \nu$)  = 0.3 (excluding Davids) or 0.7 (excluding Junghans).  Excluding $\it{both}$ Davids $\it{and}$ Junghans gives P($\chi^2, \nu$)  = 0.94, whose likelihood is similar to the P($\chi^2, \nu$)  = 0.07 fit to the full data set.  Thus, there is no reason from these fits to throw out any of the experiments~\cite{e2vse1}.

In cases like this, where results are presented from 2 very different techniques, it is important to check if the results indicate a technique dependence.  A fit to the direct data alone yields 21.4 $\pm$ 0.5 eV b, $\chi^2/\nu$ = 1.2 ($\nu$ = 4) and P($\chi^2, \nu$)  = 0.32, while a fit to the CD data alone yields 19.2 $\pm$ 0.7 eV b, $\chi^2/\nu$ = 1.0 ($\nu$ = 3) and P($\chi^2, \nu$)  = 0.39.  Thus, all the direct data are mutually consistent, as are all the CD data.  However, the probability  that these 2 mean values arise from the same parent distribution is P($\chi^2, \nu$) = 0.01, which suggests a technique dependence.  

There are other differences between CD and direct results.  JU03 presents clear evidence for a systematic difference in the slopes of S$_{17}(E)$ determined from CD data and from direct data, with probability P($\chi^2, \nu$) = 0.003 that the slopes from these 2 types of experiments arise from the same parent distribution.  In our opinion this slope difference must be understood before results of direct and CD experiments can be combined.  In addition, considerable modeling is necessary to infer $^7$Be(p,$\gamma$)$^8$B cross sections from measured $^8$B breakup data, and it is difficult to understand all the associated uncertainties (as is also the case for peripheral heavy ion transfer and breakup experiments~\cite{azhari}, which tend to yield even lower S$_{17}(0)$ values than CD experiments). 
These indirect techniques have never (to our knowledge) been ``calibrated" by comparison to a directly-measured capture reaction, with sufficient accuracy to demonstrate that they are free of systematic bias at the level of a few percent, as is necessary for indirect results to be included in determining a recommended S$_{17}$(0) value at this same level of precision.  

For these reasons, the S$_{17}(0)$ value recommended in JU03 was based on the mean of direct experiments only~\cite{centralvalue}.   It remains a challenge to understand better these differences between indirect and direct methods.

Financial support was provided by the U.S.D.O.E., grant \# DE-FG03-97ER41020.


\begin{thebibliography}{}
\bibitem{fitdifference} Fit range differences (see Table~\protect\ref{data}) are unimportant here.
\bibitem{dtparameters} DT fits with their ``$^7$Li + n" potential calculations, used for their recommended S$_{17}(0)$, yielded the same uncertainties as their ``$^7$Be + p" potential fits, central values $\sim$ 3\% lower  (see Table III of DT), and nearly identical quality fits.
\bibitem{02vs03} Including the new Seattle-TRIUMF S$_{17}$(0) value~\protect\cite{junghans03} in place of the earlier value~\protect\cite{junghans02} used by DT has a negligible effect on the fit results discussed here.
\bibitem{expterrors} Here and below we quote experimental uncertainties only, not including theoretical (extrapolation) uncertainty.
\bibitem{e2vse1} We note that the E2/E1 ratio used in interpreting the Davids data remains controversial.
\bibitem{centralvalue} The central value was based on the DB calculation, which fits the observed direct energy dependence over a wide range better than other theories, including the DT calculations.
\end{thebibliography}
\end{document}